\documentclass[10.5pt,letterpaper]{article}
\usepackage{authblk}
\begin{document}

\title{Fluid fragmentation from hospital toilets}

\date{}

\author[1,2]{G. Traverso}
\author[3]{S. Laken}
\author[4]{C.-C. Lu}
\author[5]{R. Maa}
\author[2,5]{R. Langer}
\author[6,7]{L. Bourouiba$^*$}

\affil[1]{\small MGH Division of Gastroenterology and Harvard Medical School}
\affil[2]{Koch Institute for Integrative Cancer Research, MIT}
\affil[3]{Pavoda, Inc.}
\affil[4]{ENSTA ParisTech}
\affil[5]{Department of Chemical Engineering, MIT}
\affil[6]{Department of Mathematics, MIT}
\affil[7]{Department of Civil and Environmental Engineering, MIT}

\maketitle

\let\thefootnote\relax\footnote{$^*$ corresponding author:  lbouro@mit.edu}

Hospital-acquired infections represent  significant health and financial burdens to society.  Hospital-acquired {\it Clostridium difficile} ({\it C. difficile}) is a  particularly challenging bacteria with the potential to cause severe  diarrhea and death. 
In the US alone, {\it C. difficile} is estimated to lead to more than  14,500 deaths,  and result in more than  
\$1.1 billion in costs per year \cite{Kyne2002,CDC-Cdiff,Johnston2007}. 

\medskip
One mode of transmission for {\it C. difficile}, as well as other pathogens,  which has received little attention is the potential air contamination by pathogen-bearing droplets emanating from toilets.  So far, modern mitigation strategies focused on cleaning hospital surfaces with products such as bleach, ignoring air contamination.   Though toilets have been in use for over a century,  little is known about the fluid dynamics governing their potential for pathogen-bearing droplet generation.   Best et al  (2012) \cite{Best2012} reported finding spores in the air an hour after flushing. This is a major concern given that modern hospital toilets in North America do not have lids and are cleared using high flow rate flushes.

\medskip 

Fluid dynamics is playing an increasingly important role in
understanding the transmission of a variety of diseases \cite{bb13}.  In this approach, the first step relies on characterizing the flows contributing to the pathogen transfer within or outside the infected host. 
High-speed visualizations are used to capture the  fast-time scale dynamics of fluid
fragmentation such as that occurring during real  sneezes and coughs \cite{bb12h,BDB13} or rain impacts on contaminated plant leaves \cite{GB13}. Relying on the insight gained in this first step,  complemented by clinical data, joint physical analog
experiments and  mathematical models can be developed to elucidate the fluid mechanisms contributing or shaping disease transmission  \cite{BDB13,GB13}. 

\medskip

In the fluid dynamics video submitted to the APS DFD Gallery of Fluid Motion, we present the first step of characterization of the problem of indoor contamination from hospital toilets.  Flow
visualization via high-speed recordings show the capture of the
product of the fluid fragmentation occurring during hospital
high-pressure flushing events. The recordings were performed at 1000
to 2000 frames per second and display a side view of the section just above
the toilet bowl. The visualizations capture large quantities of both large and small droplets. The former are defined as those that
would settle quickly and contaminate surfaces; while the latter could remain
suspended in the air  carrying {\it C. difficile}
spores among other pathogens \cite{Darlow1959}.  In the video, we
illustrate how high-pressure flushes and cleaning products currently used in
hospital toilets result in a dramatic increase of small droplet ejections without killing spores;
thus aggravating, rather than alleviating, the suspension and recirculation of tenacious potential pathogens. Our full results on the fluid fragmentation dynamics of flush-induced droplet ejecta will be published elsewhere. 
\medskip

\section*{Copyright Notice}
This article has been published on the ArXiv under a perpetual, non-exclusive license. Copyright is retained by the authors. The attached video files are Copyright (c) 2013 by the authors and may not be copied, publicly presented or incorporated into any other derivative work without a clear attribution and written consent.

\section*{Acknowledgments}
The authors thank H. Flexman and B. Vijaykumar for their help  with some facets of the project.

{\footnotesize

}

\end{document}